\documentclass{PoS}

\newcommand{\mni}{\noindent}

\usepackage{epsfig}
\usepackage{graphics}

\PoS{PoS(BDMH2004)}

\title{Gravitational Lensing \& Stellar Dynamics\\ {\sl Dark-Matter and
Baryons in Early-type Galaxies to z=1}}

\ShortTitle{Dark-Matter and Baryons in Early-type Lens Galaxies}

\author{\speaker{L.V.E. Koopmans}\thanks{A footnote may follow.}\\
        Kapteyn Institute, P.O. Box 800, 9700AV Groningen,
	The Netherlands\\
        E-mail: \email{koopmans@astro.rug.nl}}

\abstract{Gravitational lensing and stellar dynamics provide two
complementary, nearly orthogonal, constraints on the mass distribution
of early-type lens galaxies. This allows the luminous and dark-matter
distribution in higher-redshift ($z>0.1$) galaxies to be studied
beyond the limitations of each individual method. Two surveys have
been initiated to compile a large sample of early-type galaxies suitable to
lensing and dynamical studies: (1) The Lenses Structure \& Dynamics
(LSD) Survey and (2) the Sloan Lens ACS (SLACS) Survey . 

\mni Using spherically symmetric mass models, I illustrated how
lensing and dynamical constraints can be used to measure the
``effective'' density slope ($\gamma'$) of galaxies inside their
Einstein radii and estimate the typical error on this determination.

\mni The main results from the LSD survey thus far are: (i) Massive
(typically $>L_*$) early-type galaxies at $z$\,$\approx$\,0.5--1
contain a significant fraction $f_{\rm CDM}=0.4-0.7$ of dark matter
inside their Einstein radii. [The null-hypothesis, $f_{\rm CDM}=0$, is
excluded at the $>99\%$ in all analyzed systems.]  (ii) The inner CDM
density slope is $\gamma_{\rm CDM}=1.3^{+0.2}_{-0.3}$ (68\% CL) for
$\rho_{\rm CDM}\propto r^{-\gamma_{\rm CDM}}$. (iii) The total density
slope $\gamma' = 1.9 \pm 0.1$ (with 0.3 rms scatter in the sample;
$\rho\propto r^{-\gamma'}$).  The intrinsic scatter of 15\% in
$\gamma'$ is consistent with local dynamical studies and can
lead to a 30\% rms scatter in inferred values of H$_0$ from lens
time-delays, when purely isothermal mass models are assumed.

\mni Hence, the common practice to assume that lens galaxies are
perfectly isothermal should be abandoned, especially in cases where this
assumption is critical.
}

\FullConference{Baryons in Dark Matter Halos\\
                 5-9 October 2004\\
                 Novigrad, Croatia}

\begin{document}

\section{Introduction}

\mni Whereas the standard cosmological model ($\Lambda$CDM) is very
successful in the linear regime, in the non-linear regime observations
and theory appear to diverge. First, the inner mass profiles of dwarf
and LSB galaxies -- which are thought to be dominated by cold dark
matter -- appear to have cores with $\gamma_{\rm CDM}<1.0$
($\rho\propto r^{-\gamma_{\rm CDM}}$), much flatter than the predicted
$\gamma_{\rm CDM}$=1.0--1.5 . Second, the predicted amount of
dark-matter substructure in galaxy halos is not reflected by the
amount of visible substructure seen around the Milky Way.

\mni Although this might herald trouble for the standard cosmological
model on small scales, one `pillar' of the hierarchical structure
formation model has yet to be studied in detail beyond the local
Universe: massive early-type galaxies. Although Fundamental Plane
studies of their stellar-population evolution have been done to $z\sim
1$, comparatively little is known about their internal mass structure
and their evolution. Since early-type galaxies are predicted to form
at the highest density peaks in the early Universe, one expects them
to form with significant amounts of cold dark matter. Hence knowledge
of their dark-matter halos, their formation and their subsequent
evolution provide crucial tests of the $\Lambda$CDM model.

\mni Studying the internal mass distribution of early-type galaxies is
challenging, not only because baryonic collapse changes the initial
dark-matter mass distribution, but also because simple rotation curves
-- as for disk galaxies -- can often not be obtained and
stellar-dynamical or gravitational-lensing studies on their own suffer
from degeneracies between isotropy, mass and mass
profile. Consequently, the number of low-redshift early-type galaxies
known to have or not have dark-matter is relatively small and less is
known either about the shape of their inner dark-matter density
profiles, let alone about early-type galaxies at higher redshifts or
their structural evolution. Whereas lensing also provides some
constraints on dark-matter halos, its distribution can also often not be
measured because of the well-known mass-profile degeneracy.

\mni However, determining the amount, distribution and evolution of
baryonic (e.g.\ stellar) and dark matter in the inner $\sim$15~kpc of
early-type galaxies is a crucial step in shedding light on (i) the
$\Lambda$CDM model, (ii) the formation scenario(s) of the most massive
galaxies in the Universe {\sl and} (iii) the interaction between
baryonic and dark matter in galaxy formation (e.g.\ collisionless if
stars formed before galaxy formation, or collisional if stars formed
during galaxy formation). It is therefore essential to break
degeneracies in the existing techniques and more precisely measure the
inner mass profiles of early-type galaxies up to $z\sim 1$.

\mni The reader is referred to Kochanek, Schneider \& Wambsganss (2004)
for a thorough introduction in the above issues, that are related to
lensing.  In this proceeding, I try explain in more detail why lensing
and stellar-dynamical constraints break degeneracies in mass models
and then show how this technique is being applied to real galaxies.

\section{Gravitation Lensing \& Stellar Dynamics}

\mni\ In strong gravitational lenses (i.e.\ multiple images), the mass
enclosed by the images is nearly independent from the assumed mass
model of the lens and can often be determined to less than a few
percent accuracy. The weak dependence on higher-order moments of the
mass distribution also implies that it is difficult to measure the
mass profile of a lens galaxy, leading to the {\sl radial mass profile
degeneracy} (e.g.\ Wucknitz 2002). This degeneracy, although not
exact, is closely related to the mass-sheet degeneracy. The radial
mass profile can also be determined by comparing the kinematic
profiles (e.g.\ stellar dispersion) with those inferred from mass
models. However, the same (luminosity weighted) stellar velocity
dispersion can be the result of more a shallow/steep mass profile with
radial/tangential anisotropy of the stellar velocity ellipsoid,
especially if also the enclosed mass is allowed to vary. Hence, there
is the well-known {\sl mass--anisotropy degeneracy}.

\mni\ Because gravitational lensing determines the enclosed mass of
the lens accurately, this is precisely what is needed in stellar
dynamics to break the degeneracy with the mass profile and anisotropy
as recently shown (see Treu \& Koopmans 2004 and references therein).

\begin{figure}[t]
\begin{center}
\leavevmode
\hbox{%
\epsfclipon
\epsfxsize=0.3\hsize
\epsffile{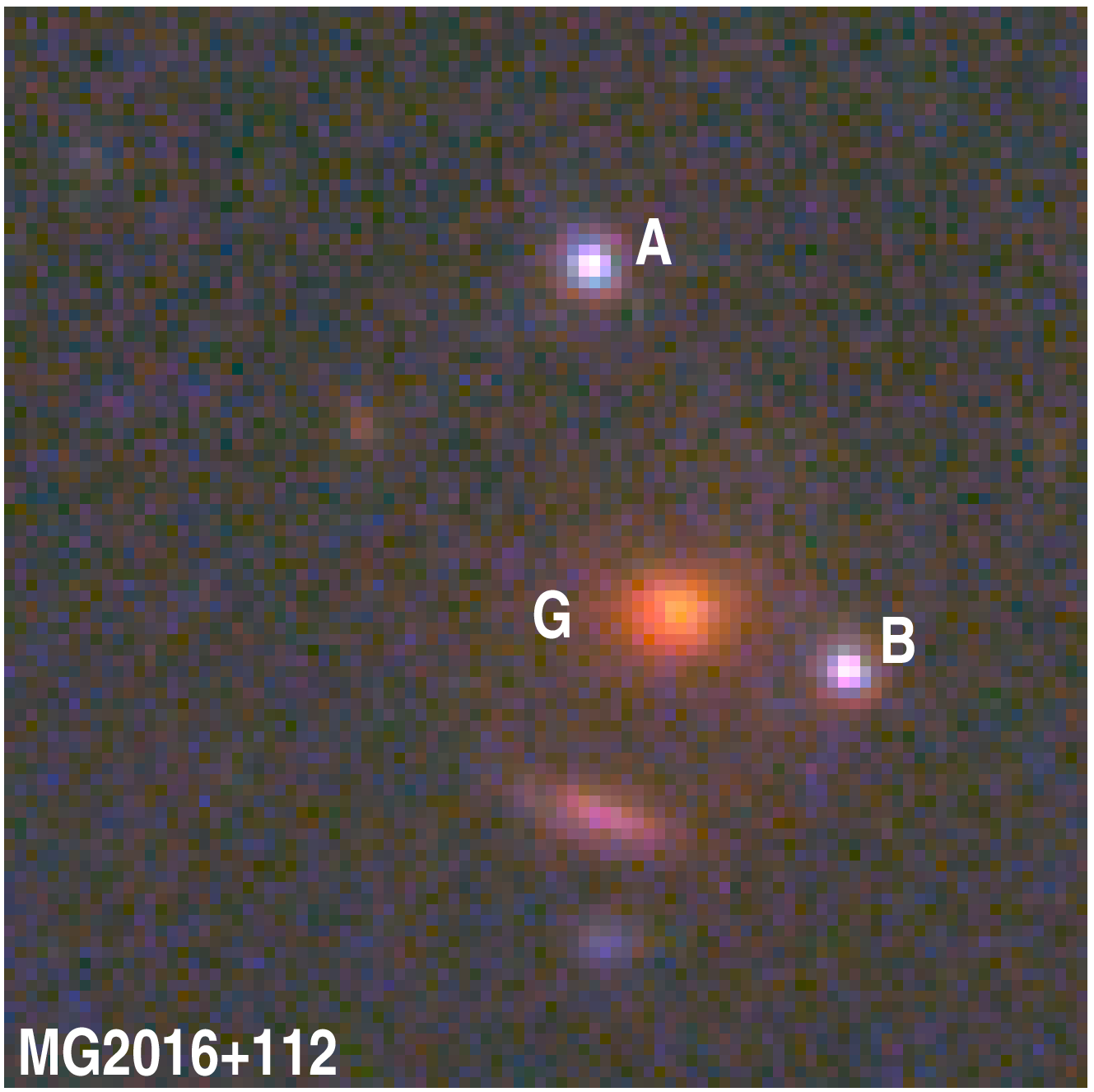}
\hspace{0.5cm}
\epsfxsize=0.63\hsize
\epsffile{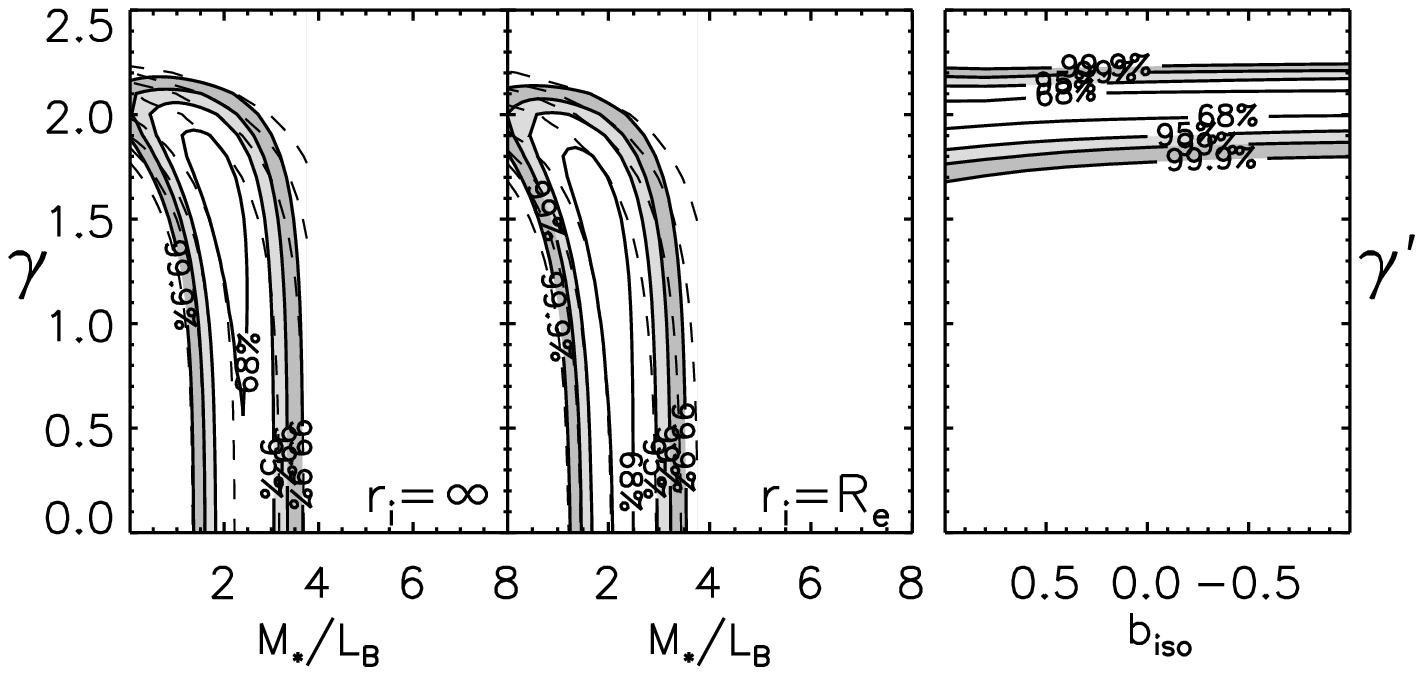}}
\epsfclipoff
\end{center}
\caption{\footnotesize {\bf Left figure:} HST true-color image of
MG2016+112 at $z$=1.004.  {\bf Right figure:} Likelihood contours in
the plane of stellar $M_*/L_B$ versus dark-matter slope ($\gamma$) for
the isotropic/anisotropic velocity ellipsoid with an Osipkov--Merritt
anisotropy radius of $r_i=\infty/R_e$.  The total $M_{\rm tot}/L_B$
inside the Einstein radius is 8.0~$M_\odot/L_{\odot,B}$, much larger
that the stellar $M_*/L_B=2.2\pm0.3$~$M_\odot/L_{\odot,B}$, indicating
an extended massive dark-matter halo. Note that the resulting total
(luminous+dark-matter) mass slopes ($\gamma'$; $\rho_{\rm tot}\propto
r^{-\gamma'}$) is nearly independent from the (constant) anisotropy
parameter $b_{\rm iso}$.}
\end{figure}

\subsection{Spherically Symmetric Lensing \& Dynamical Models}

\mni Why does the combination of stellar dynamics and gravitational lensing
work so well in determining the mass-density slope of lens galaxies?
To answer this question, the stellar velocity dispersion profile is
derived for power-law models (in density, luminosity density, etc.)
and for constant stellar velocity-ellipsoid anisotropy (i.e.\ $\beta$
models). It follows that all derived quantities then obey equally
simple power-law scaling relations. Realistic galaxies are of course
more complex. However, it serves to illustrate how different galaxies
properties are interrelated and why mass-model degeneracies can be
broken. They also allow simple error estimates of the derived quantities.

\subsubsection{Stellar Velocity Dispersion Relations}

\mni Let us suppose that the stellar component has a luminosity
density $\nu_l(r) = \nu_{l,o} r^{-\delta}$ and is a trace component
embedded in a total (i.e. luminous plus dark-matter) mass distribution
with a density $\nu_\rho(r) = \nu_{\rho,o} r^{-\gamma'}$. In addition,
let us assume that the anisotropy of the stellar component $\beta =
1-(\overline{\sigma^2_\theta}/\overline{\sigma^2_r})$ is constant with
radius. For a lens galaxy with a projected mass $M_{\rm E}$ inside the
Einstein radius $R_{\rm E}$, the luminosity weighted average
line-of-sight velocity dispersion inside an aperture $R_{\rm A}$ is
given, after solving the spherical Jeans equations, by
\begin{equation}\label{eq:sigav}
  \langle \sigma_{||}^2\rangle (\le R_{\rm A}) = \frac{1}{\pi}
  \left[ \frac{G M_{\rm E}}{R_{\rm E}} \right] f(\gamma',
  \delta,\beta) \times \left(\frac{R_{\rm A}}{R_{\rm
  E}}\right)^{2-\gamma'}
\end{equation}
with
\begin{eqnarray}
  f(\gamma', \delta,\beta) = 2 \sqrt{\pi}\,\left(\frac{\delta -3}{(\xi
  - 3)(\xi - 2\beta)} \right)&\times&
  \left\{\frac{\Gamma[(\xi-1)/2]}{\Gamma[\xi/2]} - \beta
  \frac{\Gamma[(\xi+1)/2]}{\Gamma[(\xi+2)/2]} \right\} \times
  \nonumber\\ && \left\{\frac{\Gamma[\delta/2]\Gamma[\gamma'/2]}
  {\Gamma[(\delta-1)/2]\Gamma[(\gamma'-1)/2)]}\right\}
\end{eqnarray}
with $\xi = \gamma'+\delta -2$.  Similarly,
\begin{equation}
  \sigma_{||}^2(R) = \frac{1}{\pi} \left[ \frac{G M_{\rm
  E}}{R_{\rm E}} \right] \left(\frac{\xi - 3}{\delta -3}
  \right) f(\gamma', \delta,\beta) \times \left(\frac{R}{R_{\rm
  E}}\right)^{2-\gamma'}.
\end{equation}
In the simple case of a SIS with $\gamma'=\delta=\xi=2$ and
$\beta=0$, we recover the well-known result
\begin{equation}
  \sigma_{||}^2(R) = \frac{1}{\pi} \left[ \frac{G M_{\rm
  E}}{R_{\rm E}} \right]~~~({\rm SIS}).
\end{equation}
From eqn.\ref{eq:sigav}, one sees that the radial dependence of the
stellar velocity dispersion depends on $\gamma'$ only. All other
parameters (i.e.\ $\delta$, $\beta$, etc.) only enter into the
normalization.  Since the luminosity density (i.e.\ $\delta$) is given
by the observations, as is $M_{\rm E}$ from lensing , the measurement
of $\langle \sigma_{||}^2\rangle (\le R_{\rm A} \approx R_{\rm E})$
immediately gives the density slope $\gamma'(\beta)$ (where $\beta$ in
general plays only a minor role). {\it This is the basis of combining
stellar dynamics with gravitational lensing.}

\subsubsection{Error estimate on the mass density slope}

\mni One of the most interesting parameters to determine is $\gamma'$,
i.e.\ the slope of the total mass density. We can estimate the change
$\delta \gamma'$ from the observables. One finds to first order (assuming
fixed values of $\beta$ and $\delta$):
\begin{equation}\label{eq:err1}
  \frac{\delta \sigma_{||}}{\sigma_{||}}(\le R_{\rm A}) = \frac{1}{2}
  \frac{\delta M_{\rm E}}{M_{\rm E}} + \frac{1}{2}\left(\frac{\partial
  \log f}{\partial \log \gamma'} - \gamma' \, \log\left[\frac{R_{\rm
  A}}{R_{\rm E}}\right]\right)\cdot \frac{\delta \gamma'}{\gamma'}
  \equiv \frac{1}{2} \left( \frac{\delta M_{\rm E}}{M_{\rm E}} + \alpha_g
  \frac{\delta \gamma'}{\gamma'}\right) .
\end{equation}
The second term in this equation was already derived in Treu \&
Koopmans (2002). If we further assume the errors on $M_{\rm E}$
and $\sigma_{||}$ to be independent,
\begin{equation}
  \left\langle \delta_{\gamma'}^2\right\rangle \approx \alpha_g^{-2}
  \left\{\left\langle \delta_{M_{\rm E}}^2\right\rangle + 4
  \left\langle \delta_{\sigma_{||}}^2\right\rangle \right\},
\end{equation}
where $\delta_{\dots}$ indicate fractional errors. Since in general
$\delta_{M_{\rm E}}\ll \delta_{\sigma_{||}}$, one finds the simple
rule of thumb that the error $\delta_{\gamma'} \sim
\delta_{\sigma_{||}}$ for close-to-isothermal mass models, since
$\alpha_g \sim 2$. This estimate is in very good agreement with the
results from properly solving the Jeans equations for two-component
mass models and justifies neglecting the mass errors (Treu \& Koopmans
2004). The result differs from Kochanek (2004), who uses $R_{\rm vir}$
instead of $R_{\rm A}$ and does not account for the mass term or the
first term in $\alpha_g$. Since $\sigma_{||}$ is measured
inside $R_{\rm A}\le R_{\rm E}$, not inside $R_{\rm A} \approx R_{\rm
vir} \gg R_{\rm E}$, the real error $\delta_{\gamma'}$ inside $R_{\rm
E}$ is smaller by a factor of a few than the error suggested by
Kochanek (2004), as supported by the numerical models (Treu \& Koopmans
2004).

\subsubsection{The mass-sheet degeneracy and the Hubble Constant}

\mni The mass-sheet degeneracy, $\kappa \leftrightarrow 
(1-\kappa_c)\kappa +\kappa_c$, leaves all lensing observables, except
for the time-delays (for a given value of H$_0$), invariant. The
combination of gravitational lensing and stellar dynamics provides a
way to break this degeneracy.

\mni First, we introduce two types of mass-sheets: (i) the {\sl
internal mass sheet}, i.e.\ a constant (or very extended) mass
component that effectively acts as a mass sheet {\it and} is
physically associated with the lens galaxy (thereby affecting its
stellar dynamics), and (ii) the {\sl external mass sheet}, i.e.\ a
constant mass component that is present in the direction of the lens
galaxy, but not physically associated (e.g.\ a nearly constant density
of a nearby cluster or group).  

\mni In the case of the {\sl internal mass sheet}, it is clear that
stellar dynamics breaks this lensing degeneracy. A positive mass-sheet
($\kappa_{\rm c} > 0$) lowers the average density slope inside the
Einstein radius ($R_{\rm E}$), whereas the total enclosed mass remains
the same. This lowers stellar velocity dispersion and thus a lower
density slope measured from the lensing plus dynamical models. Thus,
even though lensing can {\sl never} break the degeneracy in the true mass
slope, in combination with stellar dynamics it does.

\mni In the case of the {\sl external mass sheet}, the mass inside the
Einstein radius physically associated with the lens galaxy is
overestimated by $1/(1-\kappa_{\rm c})$. Hence, the true mass of the
galaxy is actually different than inferred from lensing by
$\delta_{M_{\rm E}}\approx -\kappa_c $. Similarly, the true mass
slope, inferred from stellar dynamics as above, will be different from
the true value by $ \delta_{\gamma'}\approx +\kappa_c
\alpha_g^{-1}$. Since the change in time-delay is $\delta_{\Delta t}
\propto (\delta_{M_{\rm E}} + \gamma'
\delta_{\gamma'}/(\gamma'-1))\propto \kappa_c
(\gamma'(\gamma'-1)^{-1}/\alpha_g -1)$ then the effects on H$_0$ tend
to be opposite for $\gamma'\approx 2$. In other words, an unknown mass
sheet (i.e.\ not accounted for in the mass model) leads to an
overestimate of H$_0$ for a fixed $\gamma'$. However, the observed
value of $\sigma_{||}$ is too small for that values of $\gamma'$ and
forces one to lower the latter, consequently lowering the inferred value of
H$_0$ again. 

\mni It should be emphasized, however, that the level of cancellation
for external mass sheets depends on the precise model and the
cancellation is neither precise nor guaranteed.

\section{Observational Programs}

\mni Currently two main programs to combine gravitational lensing and
stellar dynamical constraints are underway: (i) The Lenses Structure
and Dynamics (LSD) Survey and (ii) the Sloan Lens ACS (SLACS) Survey.
In the next two subsection, I describe each of these surveys and their
results in more details. Wheres the LSD survey has concentrated thus
far of early-type galaxies at $z>0.5$, the SLACS survey is limited to
galaxies at $z<0.5$. Their combination could provide constraint on the
evolution of the internal mass structure of early-type in the latter
half of the age of the Universe.

\subsection{The Lenses Structure \& Dynamics Survey -- LSD}

\mni The first target of the {\sl Lenses Structure \& Dynamics} LSD
survey, the lens system MG2016+112, is shown in Fig.1, as an example,
consisting of a massive elliptical galaxy (G) at $z$=1.004, multiply
imaging a quasar at $z$=3.273 in two images (A \& B). An 8.5--hr Keck
ESI spectrum was obtained from which the stellar dispersion was
determined ($\sigma=304\pm27$ km/s). Given the mass enclosed by the
images -- determined from the lens model -- one can calculate the
velocity dispersion profile of the lens galaxy, with the inner
dark-matter mass slope ($\gamma$ with $\rho_{\rm dm}\propto
r^{-\gamma}$) and stellar mass-to-light ratio ($M_*/L_B$) as free
parameters in a two-component luminous plus dark-matter mass model
(Treu \& Koopmans 2002). A useful simplification is to model the
luminous {\sl plus} dark-matter with a single density profile with
slope $\gamma'$. By comparing the model and the observed kinematic
profiles, one can place likelihood constraints on the values of
$M_*/L_B$, $\gamma_{\rm CDM}$ and $\gamma'$, as illustrated in Fig.1.

\mni We have now analyzed the dark-matter and total density profiles
of five early-type galaxies between $z \approx 0.5$ and 1.0 (Treu \&
Koopmans 2004). Based on the constraints from the current sample, we
find the following results: (i) All five early-type galaxies have
massive dark-matter halos ($>$99\% CL) with mass fractions of 40--70\%
inside their Einstein radius. (ii) The inner dark-matter slope is
$\gamma_{\rm CDM}=1.3^{+0.2}_{-0.3}$ (isotropic model) and the total
mass slope is $\langle \gamma' \rangle = 1.90$ with 0.30 rms if we
include two more analyzed systems. (iii) The stellar mass-to-light
ratio from the Fundamental Plane and from lensing \& dynamics agree
and lead to an evolution of $d \log( M/L_{\rm B})/d z$=$-0.72\pm0.10$.

\mni The results presented above are very promising, since no
limits on the dark-matter halos of early-type galaxies were known
before at this level of accuracy or at $z>0.1$. It shows that the
technique of combining lensing \& dynamics works well even
for early-type galaxies at $z\approx 1$. It also shows that early-type galaxies
exhibit a variety or spread of internal properties (i.e.\ not all
galaxies are formed equal). This emphasizes the need to study more
galaxies, covering a larger range of parameters space in
redshift, luminosity, color, etc. in order to examine possible trends,
parameter-correlations and/or evolution in more detail.

\begin{figure*}[t!]
\parbox[b]{6.5cm}{
\caption{Gravitational-lens inversion of SDSSJ1402, showing the
galaxy-subtracted HST-ACS F435W image (upper left) within a suitable
annulus.  The best-model of the system (upper right) is the mapping of
the source model (lower right) on to the image plane using the
best-fit SIE mass model. The critical curves and caustics are plotted
on the image and source models, respectively. [From Bolton et
al. 2004b]}\vspace{0.4cm}} \hfill
\resizebox{0.55\hsize}{!}{\includegraphics{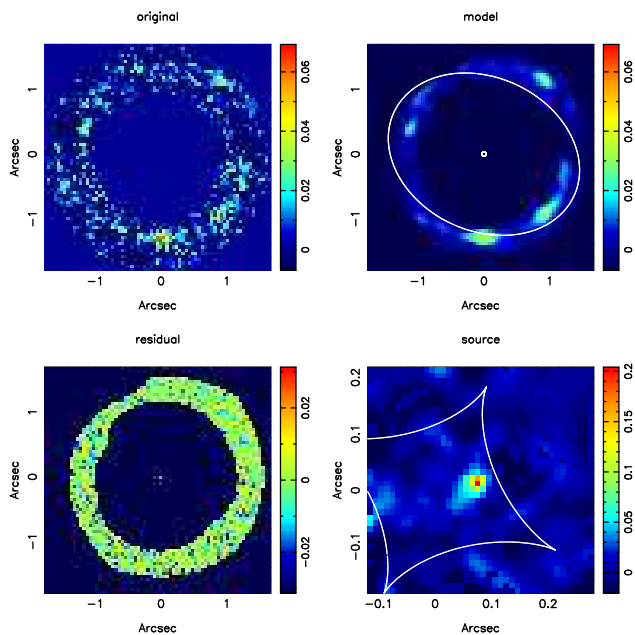}}
\end{figure*}

\subsection{The Sloan Lens ACS Survey -- SLACS}

\mni To increase the sample of lens galaxies suitable for lensing and
dynamical studies, we initiated a new HST snapshot program during
Cycle--13 (SNAP--10174; PI Koopmans): {\sl the Sloan Lens ACS Surveys
(SLACS)}. SLACS aims to target the 49 strong-lens candidates
identified by Bolton et al.\ (2004a), who analyzed $\approx$50,000
spectra from the Luminous Red Galaxy (LRG) sample from the Sloan
Digital Sky Survey (SDSS). The candidates are massive early-type
galaxies whose spectra show $\ge 3$ nebular emission lines at a
redshift higher than that of the LRG, indicating a potentially lensed
star-forming galaxy within the 3--arcsec diameter spectroscopic fiber.

\mni Bolton et al.\ (2004a) calculated that $\approx$20 of these
candidates will be strong lenses, potentially tripling the current
sample of lens systems suitable for lensing and dynamics analyzes
(including LSD lenses).  Since august 2004, one target has been
observed each week on average in the F814W and F435W filters. Thus far
about half of the sample has been targeted and $\ge$50\% are genuine
lens systems (e.g.\ Figures 2 \& 3).

\begin{figure*}[t!]
\parbox[b]{0.5\hsize}{
\caption{A newly discover SLACS lens system. A true-color image is
shown of the red lens galaxy plus a blue Einstein ring. The image was
created from two single 7-min HST--ACS images in the F814W and F435W
filters [Credit: L. Moustakas].}} \hfill
\resizebox{0.45\hsize}{!}{\includegraphics{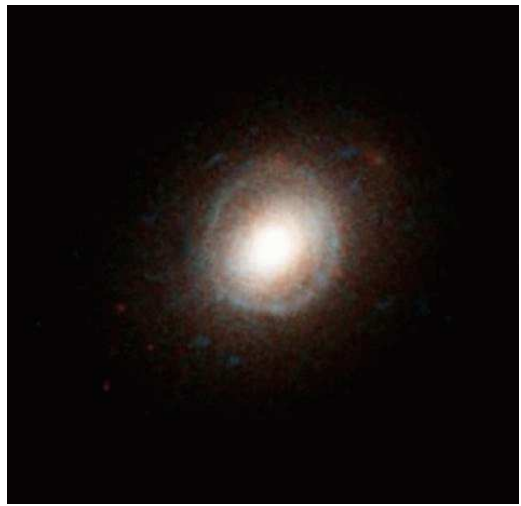}}
\end{figure*}

\mni The sample of SLACS lenses is ideal for detailed kinematic
analyzes of their lens galaxies. Because of the sample selection, the
lens galaxies are all bright LRGs and the lensed images are relatively
faint (see Figure~3). This contrasts sharply with the majority of known
lenses, where the lensed sources are bright QSOs (or galaxies) that
outshine the lens galaxy itself, complicating any kind of detailed
kinematic follow-up.

\mni To study trends and evolution of the internal structure of
early-type galaxies, we intend to measure the spatially resolved
stellar kinematics of {\sl individual} lens galaxies. Even for a small
sample this already provides extremely interesting results (see
Section~3.1), hard to obtain otherwise for galaxies beyond $z\sim
0.1$. However, a large sample of early-type galaxies with detailed
determinations of their internal structure also allows one to {\sl
compare} galaxies with different masses and as function of time, to
study trends and evolution. 

\mni In particular, with the anticipated sample of $\approx$30 lens
galaxies (including the nine observed within the LSD survey), a
detailed study of the properties and cosmic evolution of the
dark-matter halos around early-type galaxies can be made.  Assuming
the current average numbers of $\gamma'\approx 1.9$ with 0.3 rms
scatter, the $\delta_{\gamma'}$ and its rms are $15\%/\sqrt{N}$ and
$30\%/\sqrt{N/2}$, respectively, per bin with $N$ galaxies. To
distinguish changes of 10\% in $\gamma'$ at the 2--$\sigma$ level
between bins, one needs 15 objects per bin. This would allow a
measurement of the average and rms scatter {\it per bin} at the 4\%
and 10\% accuracy level, respectively.  Similarly, a total of 30
early-type lens galaxies would give the average {\it dark-matter} mass
slope to 10\% accuracy. 

\mni With an expected sample of $\sim$30 LSD\,+\,SLACS early-type lens
galaxies, changes with redshift (or mass) in $\gamma'$ at a 10\% level
can potentially be detected!

\section{Summary \& Conclusions}

\mni The combination of stellar dynamics and gravitational lensing is
starting to become a prominent tool in determining the luminous and
dark-matter mass distribution in early-type galaxies beyond the local
Universe. The reason is simple: stellar dynamics and gravitational
lensing provide unique, nearly-orthogonal, constraints on the mass
distribution of galaxies. This allows one to break degeneracies in
their mass models, even with modest imaging and spectroscopic
data-quality. 

\mni For example, the density slope ($\gamma'$) of galaxies inside the
Einstein radius ($\sim$5--15\,kpc) can be determined with an accuracy
of $\delta \gamma' \sim \delta\sigma_{||}/\sigma_{||} \sim 0.1$ (for
typical data-sets). Similarly, although less precisely, can one
determine the fraction of dark-matter ($f_{\rm CDM}$) inside the
Einstein radius and its inner slope ($\gamma_{\rm CDM}$). Results from
the LSD survey thus far give: $\gamma' = 1.9 \pm 0.1$ (with rms
intrinsic scatter of 0.3, for the LSD sample plus B1608+656 and
PG1115+080), $\gamma_{\rm CDM}=1.3^{+0.2}_{-0.3}$ (68\% CL) and
$f_{\rm CDM}(<R_{\rm E)}= 0.4-0.7$ (e.g.\ Treu \& Koopmans 2004).

\mni The SLACS survey -- a cycle--13 HST-ACS snapshot program -- is
yielding the anticipated (see Bolton et al. 2004a,b) large numbers of
lens systems at $z\le 0.5$, suitable for detailed lensing and
dynamical analyzes. The rate of discovery promises to deliver the
several dozen lens systems required to quantify the total density profile
of early-type galaxies, as well as their CDM mass fraction and
slope. 

\mni These measurement are expected to provide direct constraints
on ($\Lambda$CDM) galaxy-formation models, which properly
include dark matter, baryons and radiative processes.

\end{document}